\documentclass[manuscript]{acmart}
\AtBeginDocument{%
  \providecommand\BibTeX{{%
    \normalfont B\kern-0.5em{\scshape i\kern-0.25em b}\kern-0.8em\TeX}}}

\usepackage{multirow}
\usepackage{graphicx}
\usepackage{subcaption}
\usepackage{appendix}
\usepackage{tablefootnote}
\usepackage{dirtytalk}
\usepackage{natbib}
\usepackage{lscape}
\usepackage{pifont}
\newcommand{\cmark}{\ding{51}}
\newcommand{\xmark}{\ding{55}}

\begin{document}

\title{Islamic Lifestyle Applications: Meeting the Spiritual Needs of Modern Muslims}


\author{Mohsinul Kabir}
\email{mohsinulkabir@iut-dhaka.edu}
\orcid{0000-0001-5899-6355}
\authornotemark[]

\author{Mohammad Ridwan Kabir}
\email{ridwankabir@iut-dhaka.edu}
\authornotemark[]
\orcid{0000-0002-9631-1836}


\affiliation{%
  \institution{Islamic University of Technology}
  \streetaddress{Boardbazar}
  \city{Gazipur}
  \state{Dhaka}
  \country{Bangladesh}
  \postcode{1704}
}

\author{Riasat Siam Islam}
\email{riasat.islam@open.ac.uk}
\orcid{0000-0002-1419-8068}
\authornotemark[]

\affiliation{%
  \institution{The Open University}
  \streetaddress{Walton Hall}
  \city{Milton Keynes}
  \country{United Kingdom}}

\renewcommand{\shortauthors}{Kabir et al.}

\begin{abstract}
We evaluated contemporary Islamic lifestyle applications supporting religious practices and motivation among Muslims. We reviewed 11 popular applications using self-determination theory and the technology-as-experience framework to assess their support for motivation and affective needs. Most applications lack features that foster autonomy, competence, and relatedness. We also interviewed ten devoted Muslim application users to gain insights into their experiences and unmet needs. Our findings indicate that existing applications fall short in providing comprehensive learning, social connections, and scholar consultations. We propose design implications based on our results, including guided religious information, shareability, virtual community engagement, scholarly question-answering, and personalized reminders. We aim to inform the design of Islamic lifestyle applications that better facilitate ritual practices, benefitting application designers and Muslim communities. Our research provides valuable insights into the untapped potential for lifestyle applications to act as religious companions supporting Muslims' spiritual journey.

\end{abstract}



\maketitle

\section{Introduction}
In our research, we delve into the profound influence of religiosity on life satisfaction, a theme extensively discussed in the literature \citep{bergan2001religiosity}. In the context of Islam, we find that religious practices encompass various aspects, such as adherence to doctrines and Quranic principles, following the traditions of Prophet Muhammad, engaging in congregational worship, and contributing to charity \citep{koenig2019religiosity}. These activities are significant in pursuing spiritual enrichment among Muslims, with the ultimate goal of pleasing God and attaining inner peace in this life and divine rewards in the afterlife \citep{sholihin2022effect}. This holistic approach to Islamic lifestyle emphasizes the integration of physical, mental, and spiritual well-being, contributing to an individual's overall satisfaction with life \citep{koenig2019religiosity}.

It's worth highlighting that the impact of religiosity on life satisfaction is particularly pronounced among religious individuals, as indicated by the literature \citep{bergan2001religiosity}. Muslims worldwide actively engage in various rituals, ranging from the study of religious scriptures to acts of charity, all aimed at fostering a profound connection with the divine and achieving spiritual enrichment \citep{aziz2021analysis}. These multifaceted religious practices provide individuals with diverse means to attain a sense of spiritual fulfillment.

In the contemporary era, the widespread adoption of smartphones has significantly impacted religiosity, especially through the availability of religious applications that facilitate religious practices \cite{hameed2019survey}. Research shows that using these applications nurtures individuals' sense of belonging to their religious community, thereby contributing to increased life satisfaction \cite{bergan2001religiosity}. Moreover, motivation plays a pivotal role in sustaining religious commitment \cite{Soenens2012HowDP}. For Muslims, religious applications offer unmatched convenience, enabling them to practice their faith anywhere while swiftly accessing essential religious knowledge, thereby contributing to their inner peace \cite{hameed2019survey}. These applications typically offer features such as prayer notifications, prayer direction, access to Islamic texts, discussion forums, and tools to track religious performance \cite{hameed2019survey}. In response to the growing demand, developers continually enhance these religious applications to align with user objectives and deliver an optimal user experience \cite{rinker2016religious,kushidayati2019quran}. Factors influencing the adoption of these applications include the credibility and reliability of information, the ease and speed of accessing content, the feasibility of sustaining religious practices, and user-friendly interfaces \cite{hameed2019survey}. In summary, these applications have emerged as invaluable tools that assist Muslims in pursuing religiosity and inner peace.

Despite the growing prevalence and importance of smartphone-based religious applications, the existing literature lacks an in-depth investigation into their contribution to sustaining Islamic religiosity \cite{Neyrinck2006CognitiveAA,hameed2019survey}. To comprehensively understand the motivations driving sustained engagement in these applications, we turn to the Self-Determination Theory \cite{Ryan2000SelfdeterminationTA}. This theoretical framework provides a structured approach to exploring the psychological factors underpinning long-term application usage. Furthermore, to assess these applications' effectiveness in meeting user needs and fostering meaningful connections, we employ the Technology as Experience framework \cite{McCarthy2004TechnologyAE}. The Technology as Experience framework, with its five key dimensions — emotional, sensual, compositional, spatio-temporal, and sense-making — helps us gain a nuanced understanding of the user experience and engagement with these applications.

In our study, we comprehensively evaluate contemporary religious applications designed for the Muslim community, utilizing both the Self-Determination Theory \citep{Ryan2000SelfdeterminationTA} and the Technology as Experience framework \citep{McCarthy2004TechnologyAE}. We aim to investigate these applications' impact on user motivation and emotions. Additionally, we conduct semi-structured interviews to gain deeper insights into user engagement with these Islamic applications. The findings from our research provide valuable design implications for developers of religious applications, emphasizing the potential to sustain religious motivation by incorporating personalized and customizable features.

\section{Background}
In this section, we discuss the literature related to religiosity, its psychological aspects, motivational factors, and the integration of technology in religious practices.
\subsection{Religiosity and Life Satisfaction}
As individuals seek meaning and life satisfaction, religiosity and religious activities are pivotal in fostering psychological well-being \citep{deagon2021tools,joshi2008religious}. Spirituality and religiosity are significant influencers of well-being, with a positive connection observed regardless of religious affiliation, as they provide a sense of connection to life's deeper meaning \citep{villani2019role}. Religious commitment aids individuals in pursuing specific goals, ultimately enhancing life satisfaction. Engaging in religious activities with relevance, focus, intrinsic motivation, and a sense of fulfillment contributes to psychological well-being \citep{monson2012spiritual}.

Previous research has delved into the psychological mechanisms of rituals, highlighting that individuals, especially during times of anxiety and negativity, tend to engage in spontaneous religious rituals \citep{anastasi2008preliminary,rachman1980obsessions}. Consistent religious affiliation has been associated with numerous advantages, including social support, healthy lifestyle choices, improved self-esteem, and self-efficacy, all contributing to life satisfaction \citep{bergan2001religiosity}. Furthermore, rituals and religious practices have been shown to impact health and well-being by reducing stress and fostering social connectedness, ultimately leading to better overall health \citep{hill2010biopsychosocial}. Additionally, practices like prayer have been linked to positive mental and physical health effects by reducing depression and anxiety and promoting optimism and happiness \citep{boelens2009randomized}. Studies have established a linear relationship between religiosity and life satisfaction, with factors like theological awareness and appreciation of religious teachings influencing this connection \citep{sholihin2022effect}. Various religious activities, such as prayers, religious participation, religious cognition, obligations to religious duties, beliefs, and private practices, contribute to an individual's well-being across emotional, cognitive, and happiness dimensions, ultimately leading to increased life satisfaction \citep{sholihin2022effect}. Additionally, religious beliefs and practices have been found to enhance psychological well-being, particularly among young adults \citep{hardy2022adolescent}.

\subsection{Islamic Religiosity and the Influence of Technology}Technology has significantly impacted integrating religious activities into daily life \citep{zhao2019impact}. The coexistence and complementation of technology and religion have been acknowledged \citep{kluver2007technological}. For the Muslim community, technological support in upholding religiosity is of paramount importance \citep{tkach2016faithapps,hameed2019survey}. Islam mandates Muslims to perform prayers five times daily to remember God \citep{amjad2019prayers}. To perform prayer correctly, Muslims require knowledge of various aspects, including ablution rules, procedures, actions invalidating prayer, and Quranic chapters and verses \citep{manaf2015aplikasi, amjad2019prayers}. When traveling, they may also need essential information such as the call to prayer, direction for prayer, and mosque locations \citep{amjad2019prayers}. Additionally, Muslims often seek religious guidance, referencing the Quran and Prophet Muhammad's sayings \citep{hameed2019survey,aziz2021analysis}. The Muslim community widely utilizes smartphone-based religious applications to fulfill these needs, enabling them to acquire and disseminate Islamic knowledge, thus fostering fulfillment and sustaining religiosity regardless of their location \citep{hameed2019survey}.

In this context, smartphone applications have significantly advanced in supporting individuals' pursuit of religiosity \citep{hameed2019survey}. For example, \citet{ghembaza2018qurani} introduced "Qurani Rafiqi," an innovative Quranic application employing a context-aware and user-companion design approach. This application tracks users' time and location, retrieving and automatically notifying context-relevant Quranic verses. It also offers a voice-search feature and audio recitations by thirty-six famous reciters, eight renowned interpretations of the Quran, and four translations. These features enable smartphone users to comprehend and integrate Quranic verses into their daily lives readily. However, studies in this domain often lack analysis of fundamental questions, such as the motivations behind users' initial downloads of these applications and their overall user experiences. Moreover, these inquiries are rarely explored through the framework of theoretical models, such as Self-Determination Theory \citep{Ryan2000SelfdeterminationTA} and Technology as Experience framework\citep{McCarthy2004TechnologyAE}.

\subsection{Motivation}
Motivation plays a significant role in an individual's religiosity \citep{Soenens2012HowDP}. It provides a framework for analyzing motivations in religious practice \citep{Ryan2000SelfdeterminationTA}. Self-Determination Theory emphasizes three key components: autonomy, competence, and relatedness. Autonomy encompasses self-regulation, meaningful justifications for behavioral changes without external pressure, and consideration of other aspects of conduct. Competence involves the feeling of capability and confidence in changing behavior or achieving goals \citep{Deci2014AutonomyAN}. Relatedness pertains to attachment, respect, and understanding in interactions with others \citep{Ryan2008FacilitatingHB}. These three pillars significantly impact satisfaction and well-being \citep{Milyavskaya2011PsychologicalNM}.

Motivation toward religious practices is categorized into three primary classes: amotivation, extrinsic motivation, and intrinsic motivation \citep{Brambilla2020UnderstandingTP}. Amotivation signifies a relative absence of motivation and self-interest \citep{Gillet2010InfluenceOC}. Conversely, intrinsic motivation entails engaging in an activity purely out of joy and self-interest, devoid of external coercion, resulting in heightened performance and persistence \citep{Ryan2000SelfdeterminationTA}. In between, extrinsic motivation involves engaging in an activity due to external factors like anticipating external rewards or punishment, feelings of worry, guilt, or shame. Amotivation positively correlates with depression and low self-esteem, making individuals uncertain about the reasons behind religious practices \citep{kotera2021positive}. Autonomy and competence often underpin intrinsic motivation, with intrinsically motivated individuals exhibiting a positive attitude toward religious practices driven by enjoyment and inner peace \citep{ryan2020intrinsic}. On the other hand, extrinsic motivation often stems from external factors and can include rewards, punishments, or feelings of obligation \citep{Ryan2000SelfdeterminationTA}. The nature of these motivations can significantly impact an individual's overall well-being and satisfaction \citep{neyrinck2006cognitive}.

From a religious perspective, external factors contributing to extrinsic motivation may encompass the expectation of external rewards or punishment, feelings of worry, guilt, or shame for not engaging in an activity, recognizing an activity's importance in achieving a specific goal, self-examination followed by internalization and assimilation of reasons behind an activity \citep{Gillet2010InfluenceOC}. Research has shown that self-determined extrinsic and intrinsic motivation are significantly positively correlated with life satisfaction and self-esteem, exerting more influence on spontaneous engagement in religious practices \citep{davis2016religious}.  \citep{sheehan2018associations}. Consequently, when individuals internalize their motivation for practicing their religion, they perceive a sense of freedom and spontaneity, whereas external reinforcement is experienced when the motivation is externalized \citep{Ryan2000SelfdeterminationTA}.

\subsection{Positive Computing, Affective Interaction, and Religiosity}
"Positive computing" emphasizes technology's role in promoting good health and psychological well-being \citep{calvo2013promoting}, while "affective interaction" relates to the emotional aspects of technology interaction \citep{lottridge2011affective}. In the context of religiosity, religious applications offer these benefits. However, the diversity of these applications across genres and disciplines presents a challenge in creating a comprehensive guideline for evaluating user experiences. The Technology as Experience framework, developed by \citet{McCarthy2004TechnologyAE}, offers an innovative approach to assess user experiences across different application phases, explicitly addressing the "how" and "what" aspects of the user experience. Technology as Experience framework encompasses five components: emotional, sensual, compositional, spatio-temporal, and the sense-making process. These components collectively enhance the overall user experience.

The user's emotional state following interaction with an application serves as an indicator for evaluating the user experience. Designing applications with users' emotional experiences in mind, including sensations like thrill, anger, fear, or pain, can be crucial. For instance, a Quran recitation application with frequent sexually provocative pop-up ads \citep{kushidayati2019quran} might provoke anger and restlessness among its users, potentially leading them to discontinue use or even uninstall the application. The "sensual" component involves the immediate emotional response of a user during interaction with an application \citep{McCarthy2004TechnologyAE}. Therefore, considering users' emotional experiences, such as satisfaction and calmness, while using an application is vital. The "compositional" component evaluates the applicability of an application to its users, considering factors like user flows \citep{McCarthy2004TechnologyAE}. The "spatio-temporal" component emphasizes the importance of considering the time and place of technology interaction, as these factors significantly impact user experiences \citep{McCarthy2004TechnologyAE}. The "sense-making process" component aims to comprehend how users interpret technology while interacting with it, offering insights into their perceptions and understandings \citep{McCarthy2004TechnologyAE}.

In summary, these components collectively contribute to a holistic understanding of the user experience, especially in the context of religious applications and their impact on religiosity and emotional well-being. Despite the growing presence of Islamic lifestyle applications and their potential to facilitate religious practices among Muslims, we have identified a notable gap in the existing literature \citep{hameed2019survey, rinker2016religious}. Firstly, most studies have primarily focused on the functionalities and features of these applications, with limited attention given to the underlying motivational factors that drive our sustained engagement \citep{hameed2019survey}. While Self-Determination Theory has been employed to explore motivations in various domains \citep{Ryan2000SelfdeterminationTA}, we have found that its application within the context of Islamic lifestyle applications and how these applications foster or hinder autonomy, competence, and relatedness remains an underexplored area \citep{Ryan2008FacilitatingHB}. Secondly, we noted that the Technology as Experience framework offers a comprehensive lens for assessing user experiences, encompassing emotional, sensual, compositional, spatio-temporal, and sense-making dimensions \citep{McCarthy2004TechnologyAE}. However, there is a lack of research that employs this framework to analyze how Islamic lifestyle applications impact the emotional and sensory aspects of our religious experiences \citep{McCarthy2004TechnologyAE}, as well as our overall engagement and satisfaction \citep{Ryan2008FacilitatingHB}. Bridging these gaps is essential for us to gain deeper insights into how Islamic lifestyle applications support religious practices from motivational and experiential perspectives, ultimately informing the design of more user-centered applications for Muslim communities \citep{hameed2019survey}.

\section{Study I: Evaluating Islamic Lifestyle Applications}
   Existing literature highlights the pivotal role of motivation in religious practices and identifies a deficiency in existing applications in addressing this aspect. Therefore, Study I comprehensively evaluates popular Islamic applications that aim to assist Muslims in maintaining their religiosity. The primary objectives of this study are threefold: 
\begin{enumerate}
    \item Investigate the strategies these applications employ to leverage contextual factors in motivating users.
    \item Examine how these applications capture, record, and manage users' data and objectives, particularly in their religious pursuits.
    \item Provide an in-depth analysis of the overall user experience facilitated by these applications.
\end{enumerate}
We adopt the Self-Determination Theory as our theoretical framework to assess user motivation to achieve these goals. Additionally, we employ the Technology as Experience framework to investigate how these applications support users in adhering to their religious practices. 
\subsection{Study I: Method}
    A systematic search was carried out of the Apple App Store and the Google Play Store to create a shortlist of the applications for further analysis, the inclusion criteria for which were as follows: 
    \begin{enumerate}
        \item The applications must claim to support and notify Islamic duties in English.
        
        \item They should be freely available for download on Android and iOS operating systems.

        \item Popularity among users, as indicated by user reviews and ratings (with over 50,000 combined ratings on the Apple App Store and Google Play Store by April 2021).

        \item Categorization under the "Lifestyle" category, confirmed through search terms such as "islam + lifestyle," "muslim + lifestyle," "islam + practice," or "islam + apps."

        \item Recent development or updates to ensure ongoing technical support.

        \item Self-contained functionality, meaning they do not require third-party tools to operate.
    \end{enumerate}
    \begin{table}[htbp]
        \centering
        \caption{Functionalities of the Apps}
        \label{tab:functionalities}
        \resizebox{\linewidth}{!}{%
        \begin{tabular}{llccccccccccc}
        \hline
        \textbf{} & \textbf{Name of the app} & \textbf{PTN$^1$} & \textbf{DP$^2$} & \textbf{S$^3$} & \textbf{DQ$^4$} & \textbf{PT$^5$} & \textbf{NHR$^6$} & \textbf{NM$^7$} & \textbf{ADN$^8$} & \textbf{IGS$^9$} & \textbf{IQ$^{10}$} & \textbf{IDG$^{11}$} \\ \hline
        (1)& Athan & \cmark & \cmark & \cmark & \cmark & \xmark & \xmark & \xmark & \cmark & \cmark & \xmark & \cmark \\
        (2)& Sajda & \cmark & \cmark & \cmark & \cmark & \xmark & \xmark & \xmark & \xmark & \xmark & \xmark & \xmark \\
        (3)& AlMosaly & \cmark & \cmark & \cmark & \cmark & \cmark & \xmark & \cmark & \cmark & \xmark & \xmark & \xmark \\
        (4)& My Prayer & \cmark & \cmark & \cmark & \cmark & \xmark & \xmark & \cmark & \xmark & \xmark & \xmark & \xmark \\
        (5)& Islam 360 & \cmark & \cmark & \cmark & \cmark & \cmark & \xmark & \xmark & \xmark & \cmark & \xmark & \xmark \\
        (6)& Muslim Pro & \cmark & \cmark & \cmark & \cmark & \cmark & \cmark & \cmark & \cmark & \xmark & \xmark & \cmark \\
        (7)& Qibla Connect & \cmark & \cmark & \cmark & \xmark & \cmark & \cmark & \cmark & \cmark & \xmark & \xmark & \cmark \\
        (8)& Muslim Pocket & \cmark & \cmark & \cmark & \cmark & \cmark & \cmark & \cmark & \cmark & \cmark & \cmark & \xmark \\
        (9)& Al-Moazin Lite & \cmark & \cmark & \xmark & \xmark & \xmark & \xmark & \xmark & \xmark & \xmark & \xmark & \xmark \\
        (10)& Muslim Assistant & \cmark & \cmark & \cmark & \cmark & \xmark & \xmark & \cmark & \cmark & \cmark & \xmark & \xmark \\
        (11)& Muslim Prayer Times & \cmark & \cmark & \cmark & \cmark & \cmark & \cmark & \cmark & \cmark & \cmark & \cmark & \xmark \\ \hline
        \end{tabular}%
        }
        \footnotesize{$^1$ Prayer Time Notifications, $^2$Direction for Prayer, $^3$ Supplications, $^4$ Digital Quran, $^5$ Progress Tracker,$^6$ Nearby Halal Restaurants, $^7$ Nearby Mosques, $^8$ Arabic Days Notification,$^9$ Islamic Greetings Sharing, $^{10}$ Islamic Quiz, $^{11}$ Islamic Discussion Groups}\\
    \end{table}
    These standards were implemented to exclude low-quality applications from our analysis. Following these inclusion criteria, we identified and selected 11 applications; their specifics are shown in \autoref{tab:study_i_apps}. An overview of their functionalities is presented in \autoref{tab:functionalities}, while detailed information is provided in Section \ref{subsec:results app review}. Each application was downloaded and installed on a Xiaomi Mi A2 smartphone (running on Android v11) and on an iPhone X (running on iOS 14). Additionally, all applications were evaluated based on their free plan, with any premium or subscription-based features disregarded. Our evaluation of the selected applications encompassed the following aspects:
    \begin{enumerate}
        \item The means by which users make sense of the application. 
        \item User engagement and connection with the application.
        \item User flow for completing religious activities and the overall user experience.
        \item Mechanisms for issuing reminders and motivation prompts.
        \item Support for performing and monitoring religious practices.
        \item Features facilitating sharing of progress or information with social media or the community (Relatedness).
        \item Provision of feedback, tips, and advice to enhance religious commitment (Competence).
        \item Delivery of religious information and motivation with proper sources (Autonomy).
    \end{enumerate}
These evaluation criteria were derived from the Self-Determination Theory \citep{Ryan2000SelfdeterminationTA} and the Technology as Experience framework \citep{McCarthy2004TechnologyAE} to assess motivational and affective aspects of user interactions with the applications. To mitigate potential subjectivity, two evaluators (authors) were engaged. Both evaluators independently tested the applications over a two-week period, assigning ratings from 1 (very poor) to 5 (excellent) for each criterion and providing explanatory justifications for their ratings.

After the two-week assessment period, both evaluators convened with a mediator (author) to discuss their observations and ratings, ultimately reaching a consensus on final scores for each criterion. Subsequently, an average overall score was calculated for each application based on the components of the Self-Determination Theory and the Technology as Experience framework, as detailed in \autoref{tab:sdt and tae ratings}.
\subsection{Study I: Findings}
This section presents the findings of our investigation into how mobile applications support and motivate users in their religious practices. We evaluated these applications using the Self-Determination Theory and the Technology as Experience frameworks, focusing on their ability to promote autonomy, competence, and relatedness and their overall user experience.

\begin{table}[htbp]
        \centering
        \caption{Ratings of different applications in terms of the components of the Self Determination Theory and the Technology as Experience frameworks.}
        \label{tab:sdt and tae ratings}
        
        \resizebox{\textwidth}{!}{%
            \begin{tabular}{lccclcccccc}
                \toprule
                 \textbf{Name of the app}&
                  \multicolumn{3}{c}{\textbf{Self-Determination Theory}} &
                   &
                  \multicolumn{5}{c}{\textbf{Technology as Experience}} &
                  \multicolumn{1}{l}{\multirow{2}{*}{\textbf{Average}}} \\ \cline{2-4} \cline{6-10}
                 &
                  \textbf{Autonomy} &
                  \textbf{Competence} &
                  \textbf{Relatedness} &
                   &
                  \textbf{Sensual} &
                  \textbf{Emotional} &
                  \textbf{Compositional} &
                  \textbf{Spatio-Temporal} &
                  \textbf{Sense-Making} &
                  \multicolumn{1}{l}{} \\ \midrule
                Muslim Pro       & 3 & 4 & 5 &  & 4 & 3 & 4 & 4 & 4 & 3.88 \\ 
                Muslim Assistant & 3 & 1 & 2 &  & 3 & 2 & 3 & 3 & 4 & 2.63 \\ 
                Muslim Pocket    & 4 & 3 & 5 &  & 4 & 3 & 5 & 4 & 5 & 4.13 \\ 
                Athan            & 4 & 5 & 4 &  & 4 & 4 & 3 & 3 & 4 & 3.88 \\ 
                \begin{tabular}[c]{@{}l@{}}Muslim Prayer\\ Times\end{tabular} &
                  5 &
                  3 &
                  5 &
                   &
                  4 &
                  3 &
                  3 &
                  4 &
                  5 &
                  4.00 \\ 
                AlMosaly        & 2 & 5 & 2 &  & 3 & 2 & 1 & 2 & 2 & 2.38 \\ 
                My Prayer        & 1 & 1 & 1 &  & 2 & 2 & 1 & 2 & 3 & 1.63 \\ 
                Islam 360        & 3 & 3 & 4 &  & 4 & 5 & 3 & 4 & 4 & 3.75 \\ 
                Qibla Connect    & 3 & 4 & 4 &  & 4 & 4 & 5 & 4 & 5 & 4.13 \\ 
                Sajda            & 1 & 1 & 1 &  & 2 & 3 & 3 & 4 & 4 & 2.34 \\ 
                Al-Moazin Lite   & 1 & 2 & 1 &  & 1 & 2 & 3 & 2 & 4 & 2.00 \\ 
                \bottomrule
            \end{tabular}%
        }
    \end{table}

\subsubsection{Supporting Motivation through Self-Determination Theory}

\paragraph{Autonomy}
Autonomy involves having the information and opportunities to change one's behavior, which is crucial for religious applications to support sustained motivation and practice \citep{Deci2014AutonomyAN}. Our analysis found that most Islamic lifestyle applications focused on features rather than educational content to aid autonomy. For instance, applications lacked proper instructions for performing prayer and fasting rituals. Although some applications (Muslim Assistant, AlMosaly) had lectures or articles, they lacked context to meaningfully educate users. These applications did not emphasize developing spirituality and consistency of religious practice. Islamic applications need comprehensive guidance on religious practices, structured learning content with clear contexts, and messaging that inspires spiritual development and regular devotion over time to truly support Muslim users' autonomy. With thoughtful, user-centric design guided by autonomy, applications can become powerful tools on Muslims' spiritual journeys.

\paragraph{Competence}
The analysis showed that current applications do not sufficiently build competence, which requires engaging users in challenging growth. For instance, no applications enabled collaborative goal-setting where users can motivate one another towards shared objectives. Some applications (Athan, Sajda, Al-Moazin Lite) had activity trackers but presented minimal feedback without room for progress, falling short in nurturing efficacy and intrinsic motivation. While a few applications (Muslim Pocket, Sajda) collected user feedback, this was rarely incorporated to improve features that assist religious practice. To foster Muslim users' competence, Islamic lifestyle applications need gamification elements like goal-setting, robust analytics with personalized insights, and user-centered design that transforms feedback into enhanced features. This motivational scaffolding can help users gain mastery on their spiritual journey. applications that fail to build competence risk disengaging users from sustained devotion.

\paragraph{Relatedness}
Relatedness involves having warm, caring connections, which applications can provide through social features. Some Islamic lifestyle applications (Muslim Pro, Muslim Pocket, My Prayer, Al-Moazin Lite) integrated platforms like Facebook and Twitter to enable the sharing of festival greetings and personal updates. However, this disregards the privacy concerns of users who do not wish to share personal data on social media. For Islamic applications to fulfill relatedness, they need inclusive social spaces for mutual support beyond generic social media sharing. Small groups facilitated by scholars could enable meaningful bonds and knowledge exchange. Social rewards like progress badges shared within trusted circles could motivate users. By enabling supportive connections that align with users' privacy preferences, applications can fulfill the human need for relatedness on one's spiritual journey.

\subsubsection{Enhancing User Experience through Technology as Experience framework}
Our analysis found that applications with simpler interfaces, fewer features, and a focus on user experience better established meaningful connections per the Technology as Experience framework. For instance, Qibla Connect scored the highest by organizing features effectively and promoting usability. Despite having only four features, Sajda achieved an impressive score due to its interactive user interface, effectively establishing a meaningful connection with its user base. Meanwhile, applications like AlMosaly and My Prayer scored the lowest by failing to engage users meaningfully. 

Additionally, most applications contained advertisements that required paid subscriptions for removal. Some of these advertisements can be considered inappropriate and explicit for Muslim users. For Islamic applications, such distractions can diminish the spiritual experience. 

Islamic lifestyle applications need engaging user interfaces without intrusive advertisements or content to improve motivational impact. By carefully crafting the user experience to align with Muslim users' needs, developers can facilitate greater engagement along one's spiritual journey.
  \begin{center}
        \begin{figure}
            \begin{subfigure}{.3\textwidth}
              \centering
              \caption{Interface of \textit{Athan}}
              \label{fig:athan}
            \end{subfigure}
            \begin{subfigure}{.3\textwidth}
              \centering
              \caption{Interface of \textit{Muslim Assistant} }
              \label{fig:muslim_pocket}
            \end{subfigure}            
            \begin{subfigure}{.308\textwidth}
              \centering
              \caption{Interface of \textit{Islam 360}}
              \label{fig:vmuslim}
            \end{subfigure}
        
            \caption{Screenshots of user interface of selected Islamic applications.}
            \label{fig:screenshots}
            \Description{A three part figure, numbered from a to c in order from left to right, showing the screenshots of the home screen of user interfaces of three Islamic applications, namely - Athan, Muslim Assistant, and Islam 360. In all the three figures, there is a navigation bar at the bottom and a quick access toolbar just above the lower half of the screen, which provides users access to other features of the Applications. All the applications facilitate visualization of user's current location, prayer times, current date and time. Only in figure a (Athan application), there is an advertisement banner on top of the navigation bar. Figure c (Islam 360 application), facilitates a search bar just above the quick access toolbar for searching verses from the Quran and Hadith.
            }
        \end{figure}
    \end{center}

\subsubsection{Critical Evaluation of Key Functionalities}
\label{subsec:results app review}
This section critically evaluates the key functionalities offered by Islamic lifestyle applications.
\paragraph{Prayer time notification}
All the applications we evaluated provided prayer timings for the five daily Muslim prayers and offered customizable push notifications for the call to prayer. However, some applications, like Muslim Assistant, Muslim Prayer Times and Athan, occasionally experienced notification delays, which could be attributed to potential bugs in their software. In the cases of Qibla Connect and Al-Moazin Lite, prayer times turned out to be inaccurate due to discrepancies between user locations and the data sources for location-appropriate prayer times. Notably, Islam 360 and Qibla Connect allowed users to customize prayer times based on the four Islamic Jurisprudence schools \citep{motzki2002origins}. On the other hand, AlMosaly and Sajda encountered notification issues, with notifications sometimes failing to appear.
\paragraph{Direction for prayer}
All the applications include a feature to determine the direction for prayer based on user location data. Notably, applications like AlMosaly employ multiple methods to indicate this direction, including using a compass, reference to celestial bodies such as the Sun and the Moon, and visual representations. In contrast, Sajda and Muslim Assistant offer a live broadcast of the direction for prayer via YouTube within the application.
It is important to highlight that many of these applications rely on freely available databases for location data and corresponding prayer times. Any inaccuracies present in these databases can result in inaccuracies in determining the direction for prayer within the app. For example, according to user reviews on the Google Play Store, My Prayer and Al-Moazin Lite have been reported to provide incorrect directions. This is a significant concern, as an incorrect direction can invalidate a Muslim prayer. Additionally, user reviews on the Google Play Store for Qibla Connect have reported instances where the compass functionality becomes unresponsive.
\paragraph{Reminders}
Applications such as Muslim Pro and Muslim Pocket send reminders to users in the form of verses from the Quran and sayings of the Prophet Muhammad. It is noteworthy that, with the exceptions of Islam 360, Al-Moazin Lite, Sajda, and My Prayer, all the other applications provide reminders about significant events in the Islamic calendar, including Eid (annual Muslim festival) and Ramadan (month of fasting). Furthermore, AlMosaly enhances users' knowledge of Islamic history by sending special daily notifications labeled "On this day in history."
\paragraph{Integrated Quran}
Except for Al-Moazin Lite, all the other applications have incorporated the Quran and explanations of its verses. Users can download translations in their preferred languages to aid in comprehending the original Arabic text of the Quran. However, it is important to acknowledge that this feature sometimes encounters issues in several applications, like Muslim Prayer Times, where attempts to download translations may fail intermittently. Additionally, Qibla Connect provides a "Search Quran" feature that allows users to search for specific words within the Quran; nevertheless, this feature was found to occasionally be unresponsive.
\paragraph{Collection of Supplications}
In the context of Islam, supplication represents the act of remembering and praising God. This practice involves Muslims reciting specific Arabic phrases to seek Allah's favor, with the choice of phrases being context and time-dependent. Almost all applications offer users a wide array of supplications. However, it's noteworthy that certain applications, as indicated by user reviews, contain inaccuracies in how these supplications and remembrances are presented. These inaccuracies are concerning since they can potentially distort the intended meaning of the recited phrases. This is particularly worrisome as many Muslims may recite these phrases in Arabic without fully understanding their meaning.
\paragraph{Finding Nearby Mosques and Halal Restaurants}
Identifying nearby mosques and halal (permissible food in Islam) restaurants, particularly in regions where Muslim-minority populations reside and not all restaurants adhere to halal dietary practices, is significant to Muslim travelers. Applications such as Muslim Pro, Muslim Pocket, Muslim Prayer Times, and Qibla Connect leverage user location data to facilitate the discovery of nearby mosques and halal dining establishments. However, a prevalent issue affecting the reliability of these features is the frequent inaccuracy of location tracking. This inaccuracy is further compounded by outdated data representing non-operational or relocated mosques or restaurants. Additionally, we observed discrepancies in certain halal restaurant recommendations, where some establishments were erroneously tagged as halal. For instance, within Muslim Pro, all Nando's restaurant (popular restaurant chain) branches were categorized as halal, despite only specific branches offering halal food in the UK.
\paragraph{Progress Tracking}
Several applications offer functionality for tracking religious activities. For example, Muslim Pro includes a "Prayer Tracker," allowing Muslims to record the completion of their daily five obligatory prayers. Islam 360 can track whether an individual prayed individually or as part of a congregation. Meanwhile, Muslim Pocket and Muslim Prayer Times provide tracking features encompassing Quranic recitation, fasting, and optional prayers. AlMosaly, on the other hand, features a "Daily Goal" tracker, enabling users to establish and monitor personalized objectives. Notably, these trackers rely on manual input from users to accurately log their religious activities. However, existing literature suggests that requiring manual entry can diminish adherence and consistency in activity tracking \citep{spiller2018data}. Therefore, applications should incorporate automatic progress tracking wherever possible to enhance user engagement and continuity.
\paragraph{Sharing with the Community}
Some applications, including Muslim Assistant, Muslim Pocket, Athan, Muslim Prayer Times, and Islam 360, empower users to generate and distribute Islamic content, such as favorite Quranic verses in the form of quotes and wallpapers. This content sharing capability is often utilized during significant occasions like Ramadan and Eid, the two major Islamic festivals, allowing users to connect with friends and family through social media and messaging platforms. Muslim Pro offers a "Prayer Request" section where individuals can share their prayer requests. People typically utilize this feature to seek prayers during times of illness, bereavement, childbirth, weddings, and other significant life events. It allows users to request blessings, well wishes, and prayers from the community, fostering a sense of support and unity among its users.
\paragraph{Educational Content}
In addition to their core functionalities, some applications strive to encourage users to expand their Islamic knowledge by presenting religious content. For instance, Muslim Prayer Times incorporates a "Knowledge" section with articles and videos covering essential Islamic topics, providing users with learning opportunities. The "Inspiration" section in Muslim Pro presents motivational quotes that pique users' interest in delving deeper into the study of Islam. Muslim Pocket and Muslim Prayer Times provide a "Quiz" section, allowing users to test and enhance their Islamic knowledge. Muslim Pocket further engages users with a daily notification labeled "Today's Question," ensuring their daily participation in quizzes and enriching their Islamic knowledge. The application also offers interactive feedback by displaying the percentage of correct answers. Furthermore, Muslim Pro and Islam 360 offer comprehensive guides on Muslim pilgrimage, providing users with valuable resources for understanding and preparing for these significant religious journeys. However, it is essential to note that aside from the interactive quizzes, the remaining educational content facilitates passive learning and does not actively engage users in interactive learning experiences.

\section{Study II: Semi-Structured Interviews}
The first study unveiled various shortcomings within existing religious applications tailored for Muslims. In study II, we conducted a semi-structured interview employing a grounded theory methodology to delve deeper into these issues. This study provides insights into participants' needs and principles concerning religious practices and their engagement with Islamic lifestyle applications.

\subsection{Study II: Method}
    \subsubsection{Participants}
        We selected ten male participants who participated in interviews averaging 45 minutes each. All participants, aged between 20 and 26, fell within the young adult category and resided in [COUNTRY ANONYMIZED] during the research period. This deliberate choice stemmed from the prevalence of young adults as smartphone users in [COUNTRY ANONYMIZED], making them a valuable demographic for exploring the nexus of religiosity and technology.

        Recruitment was executed through notices posted on popular social media platforms such as Facebook. Eligibility criteria mandated a consistent and active practice of Islam for at least one year. Notably, all responders to recruitment notices were male, primarily due to cultural norms and the reluctance of practicing Muslim women to engage in extended conversations with men. A similar recruitment issue within religious communities was noted by \citet{rifat2020religion}.

        For this study, participants were not required to be users of applications from Study I; they could be users of any Islamic lifestyle application. Ethical approval from our University's Research Ethics Committee was secured, and participants provided informed consent after receiving a comprehensive briefing on the study's objectives.
    
\subsubsection{Procedure}
The study commenced with a brief demographic survey designed to glean preliminary insights into participants' demographics, encompassing their occupations, age groups, and utilization of religious applications. This questionnaire typically demanded no more than two minutes to complete. Subsequently, interview sessions were conducted to delve deeper into participants' perspectives on their religious activities. These interviews explored motivations, factors influencing sustained engagement in religious practices over time, and experiences with technology and Islamic religious applications on smartphones.

During the interviews, we adopted a grounded theory approach \citep{strauss1997grounded}, transcribing interview data to unearth recurring themes. These themes informed the development of a fresh set of questions for subsequent interviews. The semi-structured interviews allowed participants to freely express their thoughts and emotions, generating rich qualitative data. This approach offered flexibility in exploring responses in-depth and tailoring questions to emerging contextual themes. All interviews were meticulously recorded to facilitate verbatim transcription. Sample questions included queries like "How frequently do you engage in religious practices: Regularly, Occasionally, Rarely?" and "Explain the role of technology in your religious practices, how do you use them?". The interview questions are provided in Appendix \ref{appendix:app2}. Interviews were conducted online using Zoom software due to the COVID-19 lockdown.
    
\subsubsection{Data Analysis}
Prompt transcription and coding of data followed each interview to prevent any loss of valuable information. By flexibly adapting interview questions to align with emerging themes and incorporating fresh insights, data saturation was rapidly achieved, resulting in a total of ten interviews.

The interview script was meticulously designed to allow themes to emerge inductively, free from any bias toward specific theories. Grounded theory methodology \citep{strauss1997grounded} was rigorously applied to ensure that themes evolved independently of existing theories, with thematic analysis serving as the approach to identify these themes. One author meticulously handled the coding and theme analysis, while other authors scrupulously validated the results to maintain rigor. 

In addition to thematic analysis, we applied the Self-Determination Theory to further enrich our understanding of the interview data, shedding light on the participants' motivation and autonomy in their religious practices.

\subsection{Study II: Findings}
    In this section, we present our findings obtained through interview analysis. Initially, we report the outcomes of our grounded theory analysis, shedding light on our participants' beliefs, values, and motivations concerning their religious practices and how Islamic lifestyle applications support these practices. Subsequently, we delve into the findings stemming from applying the Self-Determination Theory, elucidating the fundamental motivations underpinning their religious engagement and the role played by these applications in facilitating their religious practices.

\subsubsection{Insights from Grounded Theory Analysis}

    \paragraph{The Impact of Technology on Religious Engagement}
The majority of participants reported using Islamic applications for learning purposes. Since Islamic texts are primarily in Arabic, individuals seek access to Islamic knowledge to facilitate their regular religious practices. Consequently, the availability of information and materials in their native language within these applications emerged as a fundamental reason for their usage. As one participant articulated: “I have acquired abundant knowledge regarding Islamic jurisprudence through technology, like various apps, YouTube, and Facebook. While the information may not always be accurate, these platforms have constituted a valuable resource since I lack access to scholars for guidance.” (P4)
Some participants also mentioned applications like “Hisnul Muslim,” which offers an extensive collection of prayers tailored to various emotional states such as despair, joy, and concern, effectively serving as a knowledgeable companion providing guidance when needed.

    \paragraph{Importance of User Interface}
The user interface emerged as a critical aspect that participants prioritized when selecting applications. A user-friendly interface enhances the overall experience and prolongs the use of applications. The user interface was often deemed the decisive factor when choosing various applications. One participant said, “The application I use has an aesthetically pleasing user interface featuring an efficient dashboard displaying all necessary information. Every icon is intuitive and easy to understand, simplifying navigation.” (P9)
Another participant shared a similar sentiment: “I explored various Islamic applications on the Google Playstore, installing several of them. Ultimately, I only continued using those with a satisfactory user interface and uninstalled the others.” (P2)

Conversely, an unappealing user interface resulted in users discontinuing the use of an application, as one participant noted: “I stopped using ‘Digital Quran’ due to its huge font, constant scrolling, and lack of customization options. Consequently, I struggled to read smoothly.” (P9)
Many users also preferred customizable font and user interface, as one participant described: “The old app had huge fonts, making me scroll constantly. I switched to this new app, which is super flexible. I can adjust the font size, colors, and even the background image. It is awesome!” (P9)

    \paragraph{Effectiveness of Reminders}
    
Participants regarded reminders as an effective motivational tool for religious activities. Many found push notifications from applications reminding them of various religious activities beneficial. As one participant stated: “I occasionally forget to pray due to work commitments, but receiving prayer notifications from the app makes me feel guilty for missing them. This app has become akin to a religious companion, prompting me to be more diligent in my prayers.” (P5)

In addition to daily prayer alerts, periodic notifications served as reminders for spiritually enriching religious acts, as described by one participant: “Every Friday, I receive messages like, ‘Have you prayed for the Prophet today?’ at times, it is a simple reminder during a regular afternoon, saying, ‘Have you said Alhamdulillah (All praise to Allah) today?’ These reminders help me maintain gratitude toward my Lord.” (P8)

However, some participants discontinued receiving notifications from applications, leading to reduced engagement in religious activities, as one participant explained: “Reminders are the best motivation for me. I used to receive notifications from the ‘Hisnul Muslim’ app, but since updating it, I no longer receive them. Consequently, I often forget to perform my Tasbeeh prayers at the right time.” (P3)

Conversely, push notifications with concise messages were found to be more effective in engaging participants in religious activities, as one participant stated: “‘Meaningful Salah’ sends me daily prayer alerts, such as ‘Have you read the Quran today?’ These reminders are genuinely helpful. If I do not have any classes or work after prayer, I usually recite the Quran when I receive these notifications.” (P5)
However, some participants expressed dissatisfaction with receiving the same daily notifications, finding them monotonous and less motivating.

    \paragraph{Tracking Religious Activities and Customization}

Participants revealed that tracking their daily religious activities provided short-term motivation and helped set larger goals but was ineffective in the long run. As one participant noted: “I began tracking my daily prayers, initially feeling motivated by the progress chart. However, after a few days, I could not sustain it due to time constraints.” (P3) For some participants, this tracking feature was used as a reference point to resume where they had left off, as one participant explained: “I rarely find time to read the physical Quran, so I use an app instead. The app conveniently remembers my last reading page, which is a significant help.” (P2)

    \paragraph{User-Friendly Applications and Privacy Concerns}

Participants emphasized that applications incorporating multiple useful features were considered more user-friendly, particularly for elderly users. As one participant suggested: “I would recommend getting an app with many features for my grandfather. This way, he can have everything in one place and will not need to switch between apps for different functions.” (P9)
Privacy was a top priority for participants. Many users deleted “Muslim Pro,” a popular Islamic application with approximately 10 million users, upon discovering allegations of user data being sold to third parties, even though the application did not confirm these claims. As one participant stated: “After the Muslim Pro security breach, I felt extremely insecure! I had used it for 2 years but immediately deleted it from my phone.” (P9)

Some participants who did not remove the application from their phones discontinued using many of its features, as one participant described: “I am upset that my personal information was sold without my knowledge. I briefly retained Muslim Pro to find a replacement app, but no longer use location features.” (P1)

    \paragraph{Monetization and Advertisements}
Most applications required users to upgrade to premium paid subscription schemes to access most of their features, which many participants found unaffordable. As one participant explained: “I uninstalled an app after using it for 5 days because it offered only a few features for free and required an upgrade for most functionalities. I did not want to buy something without knowing how often I would use it.” (P7)

Another participant mentioned a similar issue: “The ‘Athan’ app had a feature called ‘Community’ for chatting with other users, but it required a purchase. So, I have to pay to talk with people now!” (P2)
Some applications require users to watch advertisements before using certain features if they have not upgraded to a premium subscription. The problem with advertisements was their lack of filtration, resulting in inappropriate videos appearing before users, as one participant noted: “The concerning thing is that very indecent advertisements may pop up at any time, perhaps while you are reciting the Holy Quran. It is regrettable that they cannot filter these ads!” (P5)
Another participant shared a similar concern: “I use the Quran app to read any short note of any verse of the Quran, I need to watch an advertisement, and it is frustrating!” (P7)

    \paragraph{Desire for Shareable Features}
Participants highlighted the importance of customizable features based on user needs, as this could ensure maximum utilization of the application’s features. As one participant stated: “The ‘Salah Notifier App’ lacks a personal fasting calendar. I fast three days a month following the Arabic calendar in the app, but I have to maintain a separate note for it because the app does not offer a tracking feature.” (P3)

Participants also valued the ability to connect with other users, especially friends. Many participants expressed a desire for a simple sharing feature that would allow them to share their favorite Quranic verses or Prophet’s sayings with friends through other mediums like WhatsApp.

    \paragraph{Factors Shaping Religious Commitment}
    Parental religious education in early childhood emerged as a primary motivator toward devotion. A robust foundation of religious knowledge instilled in childhood appeared to promote enduring faith in individuals (P4). As P4 described: “I began praying with my parents at age 6, not fully grasping it then but relishing the experience. As I matured, I realized the significance of those moments in nurturing my spirituality and religious beliefs.” Early parental grounding in religious practices and principles can thus foster lifelong spirituality.
    
    Alongside parents, participants cited grandparents as active religious educators for young family members. Another childhood influence was revering historical religious figures whose stories stoked a passion for the faith (P4). As P4 recounted: “As a child, I idolized Muslim heroes like Umar Bin Khattab and Khalid Ibn Walid, aspiring to emulate them. My parents encouraged this, saying regular prayer could let me follow in their footsteps.” Venerating exemplary figures from scripture and history thus inspired and motivated religious practice from a young age.

    \paragraph{Social Influence and Encouragement}
    Many participants emphasized the importance of fruitful conversations with colleagues and friends who provided religious motivation. Such discussions influenced those uncertain about religious issues (P7) significantly. As P7 described: “I always wanted to pray regularly but worried about others’ reactions. In September 2019, a conversation with a teacher refocused me from pleasing people to pleasing the Creator.” Thus, socially-derived religious motivation through pertinent peer dialogue demonstrated a particular impact on those with religious doubts.
    
    Similarly, some participants also discussed having a religious spouse as a related factor. One participant explained: “Before marriage, my religious path had ups and downs. However, my commitment to my faith has stayed steady since getting married. I have found it easier to participate in religious activities like Islamic classes and events, largely thanks to the unwavering support of my wife, who has contributed significantly to my spiritual growth.” (P9)
    
    \paragraph{The Influence of Online Discussions}
    Online discussions about religious topics significantly influence some people. However, it is important to note that most judge online religious content mainly based on who shared it rather than its quality. One interviewee described: “In 10th grade, I started praying but was not very dedicated to other aspects of Islam. Sometimes, I would browse religious posts on Facebook. A person who used to be an atheist told his story of becoming Muslim. His accounts made me rethink religion, and I often thought about them.” (P6)
    
\paragraph{Religious Knowledge and Understanding}
Most participants agreed that developing religious understanding is essential for the motivation to engage in long-term spiritual practices. For this, the importance of gaining religious knowledge should be emphasized. As one contributor explained: “Over time, my religious motivation has grown through learning. I attended mosque gatherings where hadiths were discussed and gained valuable insights that increased my interest in Islam.” (P1)

Sources of Islamic knowledge mainly focus on the Qur’an, Islam’s most important religious text. According to the participants, learning the Qur’an positively affects all other Islamic spiritual practices. As one participant explained: “I regret not starting Arabic earlier. It would have helped me understand the Qur’an better, improving my focus during prayer. When you truly grasp something, your dedication grows stronger.” (P4)

The primary goal of gaining knowledge is to understand better the Creator, which aids people in comprehending the rules set forth by religion. One participant described: “When I started practicing Islam, I initially saw it as just a set of rules. However, as I looked at each specific rule more closely, I realized why Allah (God) had established them. This deeper understanding made it easier for me to follow them. I believe that when people follow Islam with a thorough grasp of its details, maintaining consistency in practice becomes more feasible for them.” (P10)

\paragraph{Long-Term Motivators: Unity and Support}
Some participants saw the desire to help fellow religious brothers and sisters as a long-term motivator. One individual described: “When I started practicing my faith, it was simple - praying regularly, being truthful, and aiming for paradise. However, over time, I noticed how Muslims globally faced challenges in technology and education because of oppression. Even basic items like prayer mats were imported from China, showing a lack of self-reliance in Muslim countries. This realization motivated my desire to assist Muslims facing hardships.” (P5)
This sense of unity motivated several participants to become more conscious of God. As one participant described: “In 10th grade, I started thinking about Islam, and now, my goal is to support the Muslim community. I believe they face major oppression worldwide, and by helping them, I think I can find mental peace and a deeper connection with my Creator.” (P6)

\paragraph{Inspirational Role Models}

Additionally, participants cited the inspirational role of observing how strengthening faith assisted others in overcoming struggles and motivating their religious activities (P8). As P8 recounted: “Witnessing the challenges faced by peers leading to their spiritual growth and stronger divine connection made me reflect on the life I want to lead.” Modeling these examples of overcoming adversity through strengthened religious devotion provided potent motivation to perform practices aimed at forging a deeper spiritual connection.

\paragraph{Motivation through Speech and Sermons}
One-way speeches, like sermons, can also play a vital role in motivating people toward religion. One participant recounted: “I had a teacher who was also an Imam at our local mosque. On Fridays, he gave engaging talks on overcoming social stigmas using Islamic principles. His lectures felt genuine and resonated with the audience. Because of his lectures, I started seeing Islam as a way of life rather than just a set of rules.” (P10)

\subsubsection{Insights from Self-Determination Theory}

This section delves into a comprehensive examination of the values and expectations of the study's participants, focusing on their experiences with religious applications. Through the interviews conducted with the participants, this section offers a profound understanding of the participants' perspectives, utilizing Self-Determination Theory \citep{Ryan2000SelfdeterminationTA} as a pivotal framework for analysis.

    \paragraph{Intrinsic and Extrinsic Motivation}
    
The deliberate selection criteria for participants in this study aimed at individuals demonstrating consistent and sustained engagement with Islamic practices over an extended period. The motivation behind their initial involvement in Islamic practices exhibited characteristics of extrinsic motivation, stemming either from familial influences or as a means of coping with challenging life circumstances. Commencing their religious journey during their formative years, participants did not initially articulate explicit objectives regarding their religious practices. As their religious journey progressed, participants experienced a shift towards intrinsic motivation. Participants unanimously conveyed that over the years dedicated to religious engagement and deepening their understanding of Islam, they encountered internal tranquility stemming from the performance of religious practices. This phenomenon aligns closely with the concept of intrinsic motivation as articulated by Ryan and Deci \citep{Ryan2000IntrinsicAE}. Furthermore, participants consistently recognized this intrinsic motivation as one of the most pivotal factors contributing to the enduring consistency of their religious attachment over the years.

Moreover, several participants reflected on the incentives driving their religious practices. These reflections unveiled instances where participants experienced what they perceived as miraculous outcomes through their prayers and steadfast adherence to religious activities. While religious applications served as facilitators for monitoring and providing notifications for their routine religious engagements, none of the participants indicated an overreliance on technology or religious applications as the primary source of motivation for sustaining their religious activities.
    \paragraph{Autonomy and Competence}
Information derived from the interviews underscored the contemporary trend of individuals increasingly relying on diverse digital media platforms for accessing religious videos and acquiring religious knowledge. Providing relevant information through these platforms fosters autonomy and competence \citep{Ryan2008FacilitatingHB}. In this context, religious applications emerge as valuable sources for enhancing these motivational components among end-users. While participants expressed satisfaction with the informative notifications received, such as the routine distribution of Quranic verses or Hadiths, they also voiced certain limitations. Specifically, participants noted that these applications had constraints in providing focused online learning and sustaining educational engagement. Consequently, they often resorted to alternative online resources, such as websites and blogs, to complement their pursuit of pertinent insights. It is noteworthy, however, that an overreliance on online resources for religious enlightenment raises concerns about potential misinformation, a concern acknowledged by a substantial proportion of the participants.

Another effective strategy to nurture competence is the provision of feedback based on personal informatics. Several participants reported using religious applications to track their religious activities, such as the daily five prayers and Quran recitation. While such tracking mechanisms initially provided short-term motivation, participants revealed that the applications lacked analytical features or constructive feedback mechanisms to assist them in enhancing their religious endeavors. Consequently, some participants reported a gradual loss of motivation to engage with these applications on a daily basis, attributing this waning motivation, in part, to the absence of insightful feedback.
    \paragraph{Relatedness}
    In the context of Self-Determination Theory, relatedness is synonymous with respect and understanding towards an individual's need for meaningful social connections. In the realm of religious practices, a sense of relatedness to a religious community or a group sharing similar beliefs can significantly amplify an individual's motivation to engage in religious rituals and activities.

Insights gleaned from the semi-structured interviews underscored participants' dissatisfaction with the sense of "relatedness" fostered by religious applications. The feature designated for "Discussion Groups" or "Community" was often inaccessible to most users due to its premium, subscription-based nature in most of these applications. When such features were available for free, they were frequently found to be poorly organized, resulting in diminished user interest. An alternative perspective posited by participants suggests that these applications failed to effectively facilitate meaningful religious discussions or gatherings. Many participants preferred face-to-face interactions with their religious peers over digital engagement through applications. This preference can be attributed to the absence of topic-based discussion groups or a system that connects users with similar characteristics and religious objectives, as highlighted by some participants.
\section{Discussion}
    This research explored the intersection of technology, motivation, and religious practices through two complementary studies. Study I involved a systematic review of 11 popular Islamic lifestyle applications using the Self-Determination Theory and Technology as Experience framework. Study II encompassed semi-structured interviews with 10 devoted Muslim application users to gain insights into their perspectives and needs.
    
The application review in Study I revealed that current applications are often limited in nurturing long-term motivation and affective needs. Few applications effectively supported autonomy through educational content to aid spiritual development. Applications also fell short in building competence by lacking personalized progress insights, goal-setting features, and incorporation of user feedback. Additionally, social sharing or community features were limited, constraining relatedness. Similarly, intrusive advertisements and ineffective engagement hindered user experience per the Technology as Experience framework. Thus, while applications served functional roles in religious practices, their motivational and affective impact was minimal. This highlights the need to improve application design to foster enduring intrinsic motivation and meaningfully support users’ spiritual journeys.

Study II offered qualitative insights into users’ expectations and values. Participants intrinsically valued their religious practices for the inner peace they provided but lacked conscious spiritual guidance. This suggests that applications could guide users in discovering meaningful religious goals tailored to their needs. Tracking features motivated short-term consistency but needed personalization and analytics to sustain engagement. Participants emphasized that applications could be convenient educational resources but preferred direct scholarly interactions to address complex spiritual questions. Reminders were welcomed but became ineffective if repetitive or out of context. Inappropriate advertisements were highly objectionable in religious apps. Customizable features and shareability could also enhance relatedness between users.

Integrating the findings using Self-Determination Theory principles reveals key limitations in application design. Autonomy support was minimal without structured learning content and customized goals. Competence suffered due to the absence of personalized progress tracking and feedback. Limited community features constrained relatedness. Enhancing these elements could significantly improve applications’ motivational impact. The qualitative insights also underscored the need to align application design with the sensibilities and priorities of devoted Muslim users to create a more harmonious user experience.

\subsection{Design Implications}
Our research identified several design implications and considerations for improving Islamic lifestyle applications. These insights are derived from two complementary studies and interviews with users devoted to their Islamic faith.

\paragraph{Structured Learning and Personalization}

Our research underscored the need for more structured educational content within Islamic applications, comprehensively addressing beliefs, rituals, and related topics. Users strongly desired in-depth learning opportunities, highlighting the importance of guided information. By offering users well-structured guidance covering various aspects of religion, including procedures, explanations, and benefits, applications can empower individuals to practice their faith autonomously and competently.

Furthermore, personalization emerged as a critical factor in user engagement. Adaptive learning features should be considered, tailoring content to users' knowledge levels and spiritual goals. This approach includes guided lessons on unfamiliar topics and recommendations for further enrichment. Users should be able to track their progress and receive personalized coaching to meet their spiritual objectives effectively.

\paragraph{Gamification for Intrinsic Motivation}

Gamification, when thoughtfully integrated, can enhance user engagement. Game elements such as challenges, goals, reinforcements, and rewards should align with spiritual development. For example, users could earn 'wisdom points' for learning new concepts or 'gratitude points' for acts of thankfulness. This approach fosters intrinsic motivation, driving user commitment.

\paragraph{Immersive Multimedia Content}

Participants strongly desire interactive multimedia content, including videos, podcasts, and infographics. Applications should produce original religious content optimized for digital consumption. Interactive modules, quizzes, discussions, and simulations can make learning more engaging. Emerging technologies like virtual reality can be harnessed to immerse users in spiritual experiences, stimulating their spirituality effectively.

\paragraph{Fostering Social Engagement}

Social engagement emerged as a critical factor in maintaining motivation. Sharing features allow users to share religious content with their social circles. Furthermore, virtual community features that enable users to engage in religious activities as a group can deepen their sense of community and relatedness. Making these features widely accessible avoiding subscription barriers, enhances the sense of belonging among users.

\paragraph{Connecting with Scholars and Experts}

To address complex religious inquiries, applications can facilitate interactions with scholars and spiritual experts. This feature offers valuable support in Islamic jurisprudence and contemporary issues. Users should be able to ask questions anonymously and receive personalized guidance.

\paragraph{Optimized Reminders and Respectful Atmosphere}

Optimized and personalized reminders play a crucial role in motivating users. Users value consistency in notifications, particularly regarding prayer timings. Tailoring reminders to individual needs can help users prepare for religious activities on time, enhancing their devotion. Maintaining a respectful and contemplative atmosphere is equally important, achieved by implementing robust ad-filtering systems to eliminate intrusive advertisements that disrupt the user experience.

\paragraph{Monetization Strategies with Integrity}

Non-disruptive monetization strategies, such as voluntary donations and contextually relevant product recommendations, are preferable to intrusive advertisements. Ethical, value-aligned monetization methods can sustain applications financially without compromising user experience.

By incorporating these strategies and insights, developers can enhance Islamic lifestyle applications, better-serving users' needs while respecting their values and preferences.

\subsection{Designing for faith-based communities}
Our research contributes to the Human-Computer Interaction (HCI) field by investigating the intricate relationship between technology and faith-based communities, particularly within the context of Islam. We build upon existing research related to community-based interactions and computer-mediated prayer support \citep{wolf2022spirituality}, exemplified by \citet{claisse2023keeping} research of computer-mediated prayers in Buddhism practice. While prior studies have predominantly focused on communal practices, our research sheds light on the distinctiveness of both individual and communal practices within the Islamic context. Our findings underscore the potential of lifestyle applications to effectively support both individual and communal practices among Muslims and highlight ample opportunities to develop mobile technologies to enrich these aspects of religious life.

A notable feature of our research is its timeliness, conducted amidst the backdrop of the COVID-19 pandemic lockdown. Our work aligns with similar studies investigating how technology facilitated religious practices across various faiths \citep{claisse2023keeping, wolf2022spirituality}. The resonance between our findings and these studies emphasizes the significance of community engagement, shared prayers, and the exchange of Islamic resources among users, particularly during the pandemic. This holds particular relevance considering Islam's unique blend of individual and communal practices. It is important to note that several communal practices were temporarily disrupted as mosques were closed during the lockdown period.

Within the landscape of HCI and Muslim communities, previous studies have delved into sustainability practices \citep{rifat2020religion}, exorcism rituals \citep{sultana2019witchcraft}, privacy management \citep{rifat2021purdah}, and digital sermons\citep{rifat2022putting} within Muslim communities. These inquiries tackle contemporary issues like sustainability and privacy. However, our research provides a fresh perspective by examining how mobile technologies support regular religious practices within Islam. This contribution advances the growing field of HCI concerning faith-based technologies, particularly within Muslim communities. Our findings demonstrate the existing support that Islamic lifestyle applications provide for religious practices while emphasizing the potential for substantial improvements. We elucidate these insights through the lens of the self-determination theory, revealing that bolstering motivation, autonomy, and competence can enhance technology-mediated religious experiences.

In the broader spectrum of Islamic lifestyle, halal tourism represents a related subset, sharing similarities with lifestyle Islamic applications, albeit primarily oriented toward leisure activities \citep{nahdliyah2021redesigning}. In contrast, our research directly addresses the integration of these applications into Muslims' daily routines, such as the five daily prayers. We delve into how Islamic productivity applications contribute to users' performance and habits, shedding light on the factors shaping their usage patterns. This perspective provides a nuanced understanding of how Muslims seamlessly incorporate these applications into their daily lives.

Furthermore, our findings extend into the realm of education, particularly concerning Islamic education. For instance, we note the user-centered approach in the development of Quranic Arabic learning applications \citep{mustaffa2020needs}. Our findings emphasize the pressing need for personalized educational content within mobile applications to facilitate effective learning experiences for users. These insights underscore the imperative of adhering to user-centered design principles to enhance the educational resources embedded within these applications, making them more engaging and interactive for Muslim users.

\subsection{Limitations}
Acknowledging the limitations of this research and looking toward future work, we recognize that our findings may not universally apply to all Islamic religious cultures. Islam encompasses various ideological groups, each with distinct interpretations and socio-cultural traditions. Our study focused on pious Muslims in urban settings within a specific country, and while our insights could potentially extend to regions with similar Islamic practices, further research needs to understand how individuals from diverse affiliations and spiritual orientations across various geographical and cultural contexts perceive and utilize Islamic lifestyle applications.

The global prevalence of organized religion \citep{hackett2012global} highlights the importance of designing technology sensitive to the needs of faith-based populations worldwide. In line with emerging HCI literature, which acknowledges the significance of religion in technology design \citep{buie2013spirituality, woodruff2007sabbath, wyche2009sacred, rifat2021purdah, claisse2023keeping}, we encourage future studies to explore this connection across diverse religious groups on a global scale.

Turning to the specific limitations of our research, Study I solely evaluated the features of free apps, omitting an examination of premium capabilities. Both studies had relatively small sample sizes, consisting primarily of devoted Muslim males, limiting the generalizability of gender and demographic perspectives. Participants' criteria for app selection were not predetermined. Nevertheless, our studies offer preliminary insights that can serve as a foundation for future investigations.

Moving forward, it is essential to delve into female users' perspectives, assess premium app features, and investigate applications tailored to other faiths. Exploring technological advancements, such as emotion detection during prayer, warrants further investigation. Expanding participant samples to include a more diverse range of individuals can enhance the generalizability of findings. Comparative studies encompassing various religious groups and geographical regions may reveal important cultural distinctions.

Our research sets the stage for interdisciplinary explorations into the role of technology in facilitating religiosity for motivated users worldwide while acknowledging and addressing the limitations mentioned above and avenues for future work.

\section{Conclusion}

In this paper, we aimed to make two key contributions - an application review study evaluating current Islamic lifestyle applications and a semi-structured interview study to gain user insights. Our primary objective was to investigate the effectiveness of existing religious lifestyle applications in sustaining motivation and engagement for Muslim users. We approached this inquiry by applying the Self-Determination Theory and the Technology as Experience framework in our studies.

The application review and interviews indicated deficiencies among current applications in promoting enduring motivation and engagement within the Muslim community. This can be primarily attributed to inadequate structured knowledge content, lack of personalized feedback mechanisms, and insufficient support for user autonomy in pursuing spiritual objectives.

Based on these insights, we proposed several design recommendations to enhance the motivational outcomes of religious apps. Key suggestions include providing adaptive learning features, customizable analytics aligned to personal goals, social tools to enable meaningful user connections, and optimized reminders leveraging contextual data. We believe these enhancements can nurture autonomy, competence, and relatedness to create applications that effectively motivate and support Muslim users.

Further research can build on these findings through design interventions involving application prototypes incorporating our proposed improvements. Evaluating such prototypes can help refine and validate the design implications extracted from this study. This endeavor would provide valuable assistance to developers and designers of religious applications and Muslim communities who are invested in their efficacy.

In conclusion, this research offers seminal insights into the limitations of current Islamic lifestyle applications and suggests design directions for the next generation of spiritually-oriented technologies. With thoughtful implementation, our recommendations can contribute towards creating applications that meaningfully motivate and support Muslims in pursuing devotion and self-fulfillment.

\bibliographystyle{ACM-Reference-Format} 
\begin{acks}
    The authors received no funding for this work. We thank the participants of this study for their valuable input. The authors declare no conflicting interests that could alter the outcomes of the study in any manner.
\end{acks}


\newpage

\appendix

\section{Applications Used in Study I} \label{appendix:app1}
\begin{table}[htbp]
\large
\centering
\caption{Details of the Applications Used in Study I}
\label{tab:study_i_apps}
\resizebox{\linewidth}{!}{%
\begin{tabular}{lll}
\hline
Application Name    & Google Playstore Link   & Application Version Used for Review \\ \hline
Muslim Pro          & \url{https://play.google.com/store/apps/details?id=com.bitsmedia.android.muslimpro} & 11.3.4                              \\
Muslim Assistant    & \url{https://play.google.com/store/apps/details?id=com.ninetyplus.muslim}           & 4.2.08                              \\
Muslim Pocket & \url{https://play.google.com/store/apps/details?id=com.muslim.prayertimes.qibla.app} & 1.8.4 \\
Athan               & \url{https://play.google.com/store/apps/details?id=com.athan}                       & 6.2.7                               \\
Muslim Prayer Times & \url{https://play.google.com/store/apps/details?id=com.hundred.qibla}               & 4.2.01                              \\
AlMosaly            & \url{https://play.google.com/store/apps/details?id=com.moslay}                      & 9.6.4                               \\
My Prayer           & \url{https://play.google.com/store/apps/details?id=com.mobileappsteam.myprayer}                             & 2.0.1                               \\
Islam 360           & \url{https://play.google.com/store/apps/details?id=com.islam360}                    & 4.3.5                               \\
Qibla Connect       & \url{https://play.google.com/store/apps/details?id=com.quranreading.qibladirection} & 8.1                                 \\
Sajda               & \url{https://play.google.com/store/apps/details?id=com.raimbekov.android.sajde}     & 3.7                                 \\
Al-Moazin Lite      & \url{https://play.google.com/store/apps/details?id=com.parfield.prayers.lite}       & 4.0.1132                            \\ \hline
\end{tabular}%
}
\end{table}

\section{Interview questions for Study II} \label{appendix:app2}

\begin{itemize}
    \item How frequent would you identify your religious practices to be: Regularly, Occasionally, Rarely?
    \item At which age did you consciously start religious practices? What was your motivation back then?
    \item Over the years, how did you evolve as a religious person from the perspective of goals and motivation?
    \item What is your motivation for religious practices now?
    \item What do you consider to be the most important things to be consistent in religious practices?
    \item Tell me about the role technology played or plays in your religious practices (apps, YouTube, trackers, social platforms, etc.) - how did you use them?
    \item Have you ever given up on a certain technology you tried? Why?
    \item I want you to focus on a certain smartphone application that you have used for religious purposes -
          \begin{itemize}
              \item What prompted you to choose this app?
              \item Why that app?
              \item What did you like about it? What did you not like about it?
              \item Do you think it motivates you to be consistent in your religious practices? How/Why?
              \item What particular features did you enjoy/not enjoy?
          \end{itemize}
    \item Is there anything we haven’t covered that you’d like to add?
\end{itemize}

\end{document}